# Hybrid paramagnon phonon modes at elevated temperatures in EuTiO$_3$

A. Bussmann-Holder[1], Z. Guguchia[2], J. Köhler[1], H. Keller[2], A. Shengelaya[3], and A. R. Bishop[4]

[1]Max-Planck-Institut für Festkörperforschung, Heisenbergstr. 1, D-70569 Stuttgart, Germany
[2]Physik Institut der Universität Zürich, Winterthurerstr. 190, CH-8057 Zürich, Switzerland
[3]Department of Physics, Tbilisi State University, Chavchavadze av. 3, GE-0128 Tbilisi, Georgia
[4]Los Alamos National Laboratory, Los Alamos NM 87545, USA

EuTiO$_3$ (ETO) has recently experienced an enormous revival of interest because of its possible multiferroic properties which are currently in the focus of research. Unfortunately ETO is an unlikely candidate for enlarged multifunctionality since the mode softening – typical for ferroelectrics – remains incomplete, and the antiferromagnetic properties appear at 5.5K only. However, a strong coupling between lattice and Eu spins exists and leads to the appearance of a magnon-phonon-hybrid mode at elevated temperatures as evidenced by electron paramagnetic resonance (EPR), muon spin rotation (μSR) experiments and model predictions based on a coupled spin-polarizability Hamiltonian. This novel finding supports the notion of strong magneto-dielectric (MD) effects being realized in ETO and opens new strategies in material design and technological applications.

Pacs-Index: 75.85.+t, 63.70.+h, 75.30.Ds

ETO is a prototypical cubic perovskite for T>300K where the Eu ions occupy the A sublattice in ABO$_3$, thereby leaving the B-sublattice in the d$^0$ state which is favorable for ferroelectricity. The low temperature antiferromagnetic ordering of the Eu spins and their strong coupling to the dielectric properties are clear evidence for a strong spin-lattice interaction [1 – 5] which is absent in most ABO$_3$ compounds. Nevertheless, analogies to these prototypical ferroelectrics remain, especially to SrTiO$_3$ (STO), since Sr and Eu have the same valencies, the same ionic radii and consequently the same lattice constants. In addition, both systems are incipient quantum paraelectrics, which means that the complete mode softening is suppressed by quantum fluctuations [6]. Here, an important difference arises between both since STO has a finite extrapolated ferroelectric transition temperature $T_C$=37K, whereas ETO cannot be extrapolated to any finite temperature value which makes it even less susceptible to ferroelectricity. However, a novel commonality was predicted and experimentally confirmed recently between STO and ETO, namely a rotational instability of the oxygen ion octahedra at $T_S$=282K [7, 8]. This zone boundary related phase transition is analogous to the one observed in isostructural STO at $T_S$=105K [6, 9, 10]. The refinement of high resolution x-ray analysis of powder data of ETO in space group I4/mcm taken at T=100 K yielded lattice constants of

a=5.5192(1) Å and c=7.8165(1) Å [11]. The Eu are in the 4*b* position, Ti in 4*c*, O1 in 4*a*, and O2 in 8*h* with x=0.238(1), see Fig. 1.

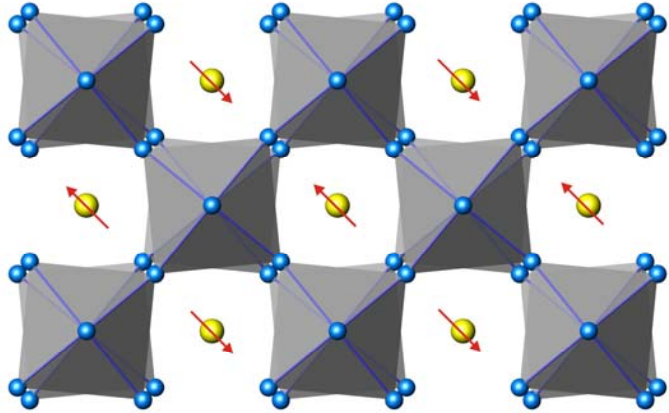

**Figure 1** Low temperature schematic structure of ETO (T=100K). For clarity the rotation angle has been enlarged by a factor of 2 and the Eu spin order below 5.5K added. The yellow circles refer to Eu, the blue ones to O and the Ti atoms are centering the octahedron being located below the apical oxygen ions.

In analogy to STO this phase transition has been related to the softening of a transverse zone boundary acoustic mode frequency which is predicted to display a very similar temperature dependence as observed in STO [7, 8] and reminiscent of a purely displacive transition. However, the calculations indicate that the related double-well potentials differ grossly: STO has a broad and shallow double-well potential whereas ETO exhibits a deep and narrow one [12]. The crossover in the dynamics between these two extremes was demonstrated by investigating the phase diagram of the mixed crystal series $Sr_{1-x}Eu_xTiO_3$, where a nonlinear dependence of $T_S$ on x was observed [12].

Here, we concentrate on the pure $EuTiO_3$ system and investigate its magnetic properties in relation to the soft mode dynamics. Motivated by the strong spin-phonon coupling observed at the onset of AFM order, similar strong interactions are expected to appear at the structural phase transition for the following reasons: From first-principles GGA+U calculations [8] it appeared that two competing interactions are present in $EuTiO_3$, namely the nearest neighbor AFM interaction $J_{nn}$ and the second nearest neighbor ferromagnetic interaction $J_{nnn}$. Both are almost of the same order of magnitude. While $J_{nn}$ should be unaffected by the oxygen octahedral rotation, $J_{nnn}$ varies with it since this is the indirect one via the bridging (and rotating) oxygen ion. This suggests that $J_{nnn}$ adopts a temperature dependence analogous to the soft zone boundary mode. In addition correlated spin fluctuations appear which are evidenced by μSR experiments.

The system is modeled within a spin-phonon coupled approach [7, 8, 13] with the phonon subsystem described by the nonlinear polarizability model [14 – 16]. This guarantees that the optic mode softening is correctly reproduced and allows a self-consistent derivation of the local double-well potential [8, 12]. In addition, predictions for the zone boundary acoustic mode softening have been made [12]. The essential ingredients of the model are the nonlinear polarizability of the oxygen ion $O^{2-}$ which is unstable as a free ion and partially stabilized by the Madelung potential of the surrounding lattice [17]. This property is modeled by an attractive harmonic core-shell coupling $g_2$ and an anharmonic fourth order coupling $g_4$ in the relative core-shell displacement coordinate $w$ where both quantities have to be derived self-consistently. The stability of the system is guaranteed by a second nearest neighbor harmonic coupling $f'$ between the polarizable units. The nearest neighbor coupling $f$ between the rigid ion sublattice and the polarizbale units together with the core-shell coupling $g_T = g_2 + 3g_4 \langle w^2 \rangle_T$ ensures mode-mode coupling and produces anomalies in the elastic constants. The coupling between the spins and the lattice, $\varepsilon$, modifies, through the lattice dynamics, the $xy$ components of the $g$ tensors, whereby $\varepsilon$ varies linearly with the magnetic field $H$. The dispersion relations for the coupled mode system have been derived in Refs. 7, 8.

For small spin-lattice coupling the zero momentum optic mode softens with decreasing temperature. In this limit the soft optic mode has the same temperature dependence as in the uncoupled case. For increasing spin-phonon coupling, respectively increasing field strengths, $\varepsilon$, ~ $H$, the soft mode frequency hardens with increasing coupling in agreement with experimental data [1 – 3]. In addition, the mean value of the z-component of the spin is affected and will depress the dielectric constant differently for fields parallel or perpendicular to it. Besides the anomalous low temperature behavior of the dielectric constant a strong coupling of the Eu spins with the optic and acoustic branch sets in with finite $\varepsilon$ which modifies the dispersion of all modes and admits for short range magnetic order above $T_N$, namely already in the paramagnetic phase. The theoretical situation for different magnon energies $\omega_0$ and with momentum q along (100) where the soft optic mode is observed, is shown in Fig. 2 as a function of temperature.

With decreasing temperature an increased magneto-acoustic coupling sets in for both magnon energies which leads to a substantial suppression of the acoustic mode for large magnon energies (Fig. 2a) at the zone boundary. This acoustic-paramagnon coupling should also be evident in the piezo-magnetic response. The optic mode is not affected by the spin mode at the zone boundary where it adopts its rigid ion value. However, at small momentum the optic mode softening gets pinned at the magnon energy with decreasing temperature and a polar instability is inhibited. Finite momentum optic mode magnon coupling exists for $q < 0.2$ with the wave vector providing information on the real space spin modulations. For $q \approx 0.2$ this corresponds to roughly 5 lattice constants. A rather similar paramagnon-phonon coupling as described above, has been observed in hexagonal $YMnO_3$ [17, 18], where 10K above the magnetic ordering temperature a spin wave mode has been observed which strongly interacts with the optic and acoustic phonon mode branches.

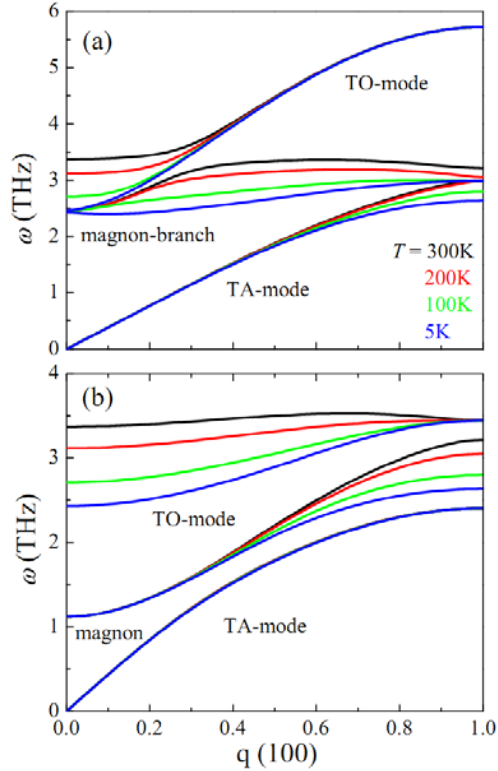

**Figure 2** Temperature dependence of the dispersion of the magnon and optic and acoustic mode frequencies for a) magnon with energy $\omega_0$=2.5 THz and b) $\omega_0$=1.2 THz. The temperatures are given by the color code shown in the figure. The calculations refer to momentum q along (100).

Such a coupling is reminiscent of the hybridized soft mode of TbMnO$_3$ [19] observed by inelastic neutron scattering. It depends, however, on the value of the paramagnon energy $\omega_0$.

For small values of $\omega_0$ (Fig. 2b), a crossing of magnon and acoustic branch takes place at small momentum leading to pronounced magneto-electric coupling. This latter evolution with temperature has been reported in Ref. 20 for hexagonal YMnO$_3$ where far above the Néel temperature short range magnetic correlations enable the observation of phonon-paramagnon coupling, rather analogous to the above results. In this case the long wave length optic mode softens similarly to the uncoupled case. At intermediate momentum values, however, the softening remains and is attributed to another crossing of the magnon branch with the optic mode.

The predicted finite size phonon-paramagnon coupling at temperatures T>T$_N$ and T$_S$, has been tested experimentally by μSR measurements. Zero-field (ZF) μSR experiments were performed at the μE1 and πM3 beamlines of the Paul Scherrer Institute (Villigen, Switzerland). The polycrystalline ETO sample has been prepared as described in Ref. 7. The sample was mounted on a sample holder with a standard veto setup providing essentially a low-background μSR signal.

In a μSR experiment nearly 100 % spin-polarized muons are implanted into the sample one at a time. The positively charged muons $\mu^+$ thermalize at interstitial lattice sites, where they act as magnetic microprobes. In a magnetic material, the muon spin precesses in the local magnetic field $B_\mu$ at the muon site with the Larmor frequency $\nu_\mu = \gamma_\mu/(2\pi)B_\mu$ (muon gyromagnetic ratio $\gamma_\mu/(2\pi) = 135.5$ MHz/T). ZF μSR is a very powerful tool to investigate microscopic magnetic properties of solids without applying an external magnetic field. A ZF μSR time spectrum for the polycrystalline $EuTiO_3$ sample recorded at 1.6 K is shown in the inset of Fig. 3a. At this temperature a spontaneous muon spin-precession is observed indicating a well-defined internal magnetic field at the muon sites, consistent with the low temperature AFM phase. The ZF μSR data below $T_N$ are analyzed using the functional form: $A(t) = A_s(t) + A_{BG}(t)$ with the first component describing the sample response and the second representing the background contribution. The sample contribution is described by the expression: $A_s(t) = A_0\left[\frac{2}{3}\exp(-\lambda_T)\cos(\gamma_\mu B_\mu t + \varphi) + \frac{1}{3}\exp(-\lambda_L t)\right]$ where $A_0$ denotes the initial asymmetry, and $\varphi$ is the initial phase of the muon-spin ensemble. $B_\mu$ represents the internal magnetic field at the muon site, and the depolarization rates $\lambda_T$ and $\lambda_L$ characterize the damping of the oscillating and non-oscillating part of the μSR signal. The 2/3 oscillating and the 1/3 non-oscillating μSR signal fractions originate from the spatial averaging in powder samples, where 2/3 of the magnetic field components are perpendicular to the muon spin and cause a precession, while 1/3 of field components are parallel and do not contribute. The extracted internal field $B_\mu$ is given in Fig. 3a, and vanishes – as expected at $T_N$. In Fig. 3b) the relaxation rates $\lambda_L$ and $\lambda_T$ are shown as a function of temperature. The transition to the AFM state is marked by a divergence in $\lambda_L$ and a peak in $\lambda_T$ (critical slowing down) upon approaching $T_N$.

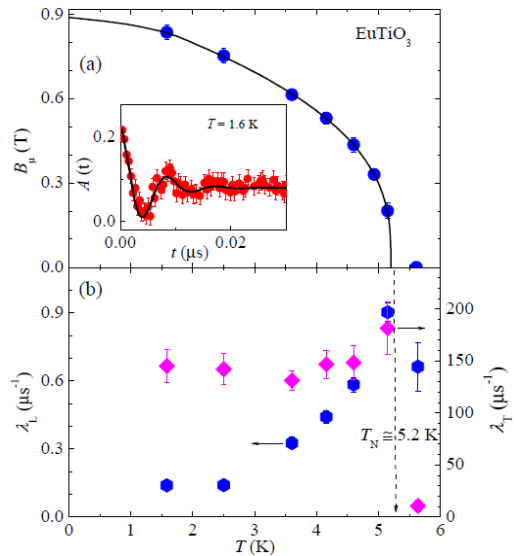

**Figure 3** a) Temperature dependence of the internal field $B_\mu$ of polycrystalline EuTiO$_3$. The inset shows the ZF µSR spectrum of polycrystalline EuTiO$_3$ below T$_N$ at T=1.6K. b) Temperature dependence of $\lambda_L$ and $\lambda_T$.

The above data evidence, that our method is very sensitive in detecting any kind of magnetic order in EuTiO$_3$.

For temperatures T>T$_N$ the oscillatory time evolution of the asymmetry vanishes, but still a damped decay rate of $A(t)$ remains, stemming from thermally induced magnetic disorder of randomly oriented spins. In this temperature range the analysis becomes statistically compatible with the single exponential component $A(t) = A_0 \exp(-\lambda_{para} t)$ with $A_0$ denoting the initial asymmetry and $\lambda_{para}$ is the relaxation rate referring to the magnetic moments surrounding the muon. The temperature dependence of the relaxation rate $\lambda_{para}$ of polycrystalline EuTiO$_3$ in the paramagnetic phase is shown in Fig. 4a. At the structural transition T$_S$ a pronounced anomaly in $\lambda_{para}(T)$ is observed, demonstrating that at T$_S$ the magnetic moments, surrounding the muon spin, change due to a change in the structure. As has been outlined above, two competing interactions between the spins exist, the nearest neighbor AFM exchange $J_{nn}$ and the next nearest neighbor ferromagnetic superexchange $J_{nnn}$ via the in between lying oxygen ions. Both interactions are closely balanced [8]. Since at T$_S$ the oxygen ion octahedra rotate anticlockwise with respect to each other [11], $J_{nnn}$ is altered at T$_S$ and induces a change in the muon spin relaxation rate caused by pronounced spin-lattice interaction. This survives even above the structural transition temperature. We demonstrate this conclusion by comparing the temperature dependence of $\lambda_{para}$ with the one of the zone boundary soft mode frequency and the EPR line width of Ref. 12 (Figs. 4b, 4c), already presented in Ref. 12. In contrast to Ref. 12 we have plotted the squared frequency $\omega_{TA}^2(q=2\pi/a)$ as a function of (T-T$_S$) and applied the standard mean-field Curie-Weiss law to the mode in the low temperature regime. In order to highlight this analogy further the mode has been shifted upwards by 3.2 THz$^2$, but actually is zero at T$_S$.

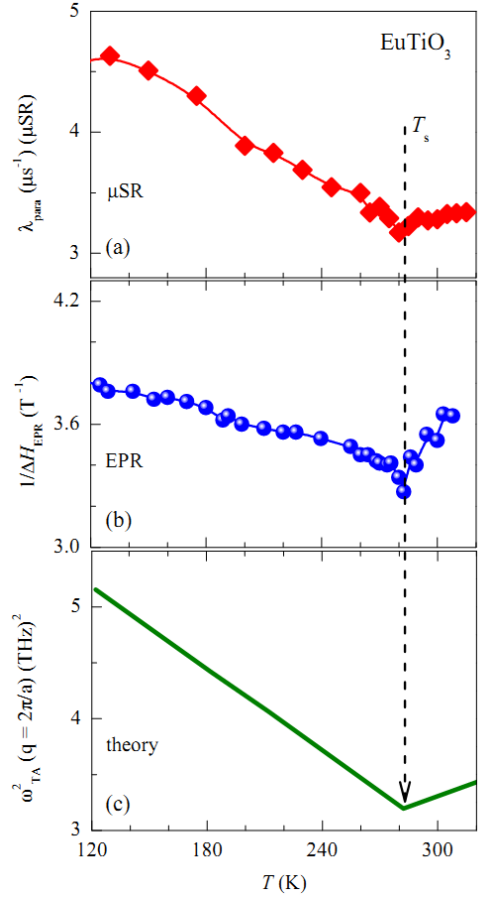

**Figure 4** a) Paramagnetic relaxation rate $\lambda_{para}$ of polycrystalline EuTiO$_3$ as a function of temperature. At the structural transition temperature T$_S$ an anomaly appears which is indicated by the arrow. The solid lines are guides to the eye. b) The EPR line width of ETO as a function of temperature [12]. c) Calculated temperature dependence of the squared soft zone boundary transverse acoustic mode $\omega_{TA}^2(q = 2\pi/a)$ along the projected (111) direction. The mode has been shifted upwards by 3.2 THz$^2$ in order to illustrate the close analogy between both figures.

While this comparison is not quantified with respect to the direct involvement of the soft zone boundary in the spin dynamics, the similarity between all three temperature dependencies is striking and demonstrates that a strong coupling between the spin and the lattice must be present. Also, it is important to note that the soft zone boundary mode shows an almost second order type phase transition which is accompanied by the slowing down of the relaxation rate at T$_S$. The analogous temperature dependence of the inverse EPR line width (Fig. 4b) is an additional support for the suggested strong spin-lattice interaction.

The relation between EPR and μSR has been derived in Ref. 22. We follow closely this analysis in the following. The temperature dependence of $1/(\Delta H_{EPR})$ shown in Fig. 4b is similar to $\lambda_{para}(T)$. In order to understand this similarity the relation between $\lambda_{para}(T)$ and the fluctuation time of paramagnetic spins needs to be considered. In Ref. 21 the depolarization rate $\lambda_{para}(T)$ measured in ZF μSR experiments has been analyzed in the paramagnetic phase of compounds containing 4f-shell ions. It was found that $\lambda_{para}(T)$ measured at temperatures high with respect to the magnetic phase transition temperature is a function of fluctuation the time $\tau$ of the 4f spins. In cubic symmetry the $\lambda_{para}(T)$ can be written as [22]: $\lambda_{para}(T) = 2\Delta_e^2 \tau(T)$, where $\Delta_e$ is where is the hyperfine coupling constant between the muon spin and the localized 4f moments. From EPR it was observed that the dominant contribution to the relaxation mechanism of the 4f ions stems from the spin-phonon coupling. Consequently, the time $\tau$ can be replaced by the spin-lattice relaxation (SLR) time $T_1$. The SLR time is defined by expression [21] $T_1(s) = 1/(7.62 \times 10^6 g \Delta H_{EPR})$ which, when combined with the above relation yields $\lambda_{para}(T) = 2\Delta_e^2(T)/(7.62 \times 10^6 g \Delta H_{EPR})$. This agrees well with the experimentally observed correlation between $\lambda_{para}(T)$ and $1/(\Delta H_{EPR})$.

Since it is well known that the actual spin ordering temperature is $T_N$=5.5K, we conclude from the data and the theoretical analysis that the Eu spins follow the lattice dynamics within spatially limited regions. They are dragged by the mode softening and fluctuate locally in an ordered manner thus giving rise to the μSR response at elevated temperatures. This finite size coupling between the spins and the optic and acoustic mode branches are expected to give rise to novel piezo-magnetic, opto-magnetic, and magneto-elastic effects. Since the spin ordering is not coherent on the lattice the system represents an inherently inhomogeneous state with locally confined dynamical interactions underlining our conclusion about the strong hybrid paramagnon-phonon coupling.

**Acknowledgement**
The experiment was performed at the Swiss Muon Source, Paul Scherrer Institute (PSI), Villigen, Switzerland. The μSR time spectra have been analyzed using the free software package musrfit [22] mainly developed by A. Suter at PSI. We thank A. Amato, Z. Shermadini, and A. Maisuradze for the support during the μSR experiments. This work was partly supported by the Swiss National Science Foundation and the SCOPES grant No. IZ73Z0_128242.